\documentclass[11pt,twoside]{article}
\usepackage{./asp2014}
\usepackage{rotating}

\aspSuppressVolSlug
\resetcounters

\begin{document}

\title{Extragalactic Proper Motions:  Gravitational Waves and Cosmology}
\author{Jeremy Darling$^1$, Alexandra Truebenbach$^1$, and Jennie Paine$^1$}
\affil{$^1$Center for Astrophysics and Space Astronomy, Department of Astrophysical and Planetary Sciences,
                 University of Colorado, 389 UCB, Boulder, CO 80309-0389, USA}%; \email{jeremy.darling@colorado.edu}; \email{alexandra.truebenbach@colorado.edu}} 
%\affil{$^2$Center for Astrophysics and Space Astronomy, Department of Astrophysical and Planetary Sciences,
%                 University of Colorado, 389 UCB, Boulder, CO 80309-0389, USA; \email{alexandra.truebenbach@colorado.edu}} 

% This section is for ADS Processing.  There must be one line per author.
\paperauthor{Jeremy Darling}{jeremy.darling@colorado.edu}{0000-0003-2511-2060}{University of Colorado}{Center for Astrophysics and Space Astronomy, Department of Astrophysical and Planetary Sciences}{Boulder}{CO}{80309-0389}{USA}
\paperauthor{Alexandra Truebenbach}{alexandra.truebenbach@colorado.edu}{}{University of Colorado}{Center for Astrophysics and Space Astronomy, Department of Astrophysical and Planetary Sciences}{Boulder}{CO}{80309-0389}{USA}
\paperauthor{Jennie Paine}{jennie.paine@colorado.edu}{}{University of Colorado}{Center for Astrophysics and Space Astronomy, Department of Astrophysical and Planetary Sciences}{Boulder}{CO}{80309-0389}{USA}

\begin{abstract}
Extragalactic proper motions can reveal a variety of cosmological and local phenomena over a range of angular scales.  These include 
observer-induced proper motions, such as the secular aberration drift caused by the solar acceleration about the Galactic Center and
a secular extragalactic parallax resulting from our motion with respect to the cosmic microwave background rest frame.  Cosmological 
effects include anisotropic expansion, transverse peculiar velocities induced by large scale structure, and the real-time evolution of the 
baryon acoustic oscillation.  Long-period gravitational waves can deflect light rays, producing an apparent quadrupolar proper motion
signal.  We review these effects, their imprints on global correlated extragalactic proper motions, their expected amplitudes, the current best measurements
(if any), and predictions for {\it Gaia}.  Finally, we describe a possible long-baseline ngVLA program to measure or constrain these proper motion signals.  
In most cases, the ngVLA can surpass or complement the expected end-of-mission performance of the {\it Gaia} mission.
\end{abstract}

\section{Background}

The universe is dynamic, as we know from cosmological redshifts induced by the Hubble expansion, but astronomers tend to treat extragalactic
objects as fixed in the sky with fixed apparent velocities.  If measured with enough precision, however, nothing is constant:  all objects
will change their redshifts and positions at a rate of order $H_0 \simeq 7\times10^{-11}$~yr$^{-1} \simeq 15$~$\mu$arcsec~yr$^{-1}$.  The secular redshift drift caused by a non-constant expansion \citep{sandage1962} is of order 0.3~cm~s$^{-1}$~yr$^{-1}$ at $z\simeq1$ and may be measured by optical telescopes
using the Ly$\alpha$ forest or by radio telescopes using \ion{H}{i} 21 cm or molecular absorption lines \citep{loeb1998,darling2012}.  Proper motions of 
extragalactic objects may be caused by peculiar motions induced by large scale structure 
\citep{darling2013,darling2018b,truebenbach2018}, 
by primordial gravitational waves \citep[e.g.,][]{pyne1996,gwinn1997,book2011,darling2018}, 
by the recession of fixed objects such as the baryon acoustic oscillation, or by anisotropic expansion 
\citep{fontanini2009,quercellini2009,titov2009,darling2014}.  

Observer-induced proper motions are also possible:  these include the secular aberration drift caused by acceleration of the solar system barycenter about the Galactic Center \citep[e.g.,][]{bastian1995,eubanks1995}, observations from a rotating reference frame, and secular extragalactic parallax caused by motion with respect to the cosmic microwave background \citep[CMB;][]{ding2009}.

Proper motions depict a discretely sampled vector field on the celestial sphere.  In order to detect and characterize correlated motions, it 
is natural to describe vector fields using vector spherical harmonics (VSH), which are the vector equivalent of the scalar spherical harmonics used
to describe signals such as the CMB temperature pattern, the geoid, or equipotentials \citep{mignard2012}.  
VSH are characterized by their degree $\ell$ and 
order $m$, and resemble electromagnetic fields.  They can therefore be separated into curl-free (E-mode) and divergenceless (B-mode) 
vector fields that are typically connected to distinct physical phenomena.  The general method for characterizing a correlated proper motion 
field is to fit VSH to the observed proper motions and to calculate the power in and significance of each degree $\ell$ for each mode.

\section{Expected (and Possible) Signals}\label{sec:signals}

Table \ref{Tab:Summary} summarizes the expected and possible global extragalactic proper motion signals.  This summary is likely to be 
incomplete.  Here we provide a brief description of each physical or observer-induced effect, and Section \ref{sec:ngvla} discusses the 
impact the ngVLA might have on detecting or constraining these phenomena.

\subsection{Secular Aberration Drift}

\begin{figure}[ht!]
%\epsscale{1.18}
\plotone{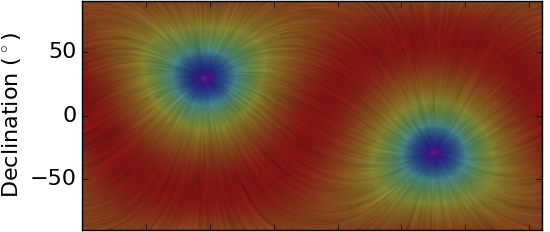}
\plotone{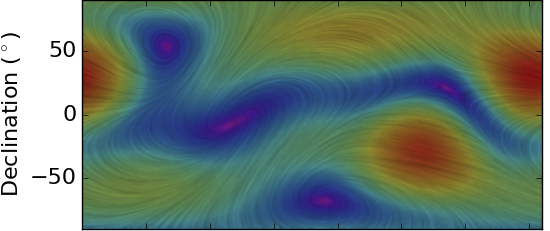}
\plotone{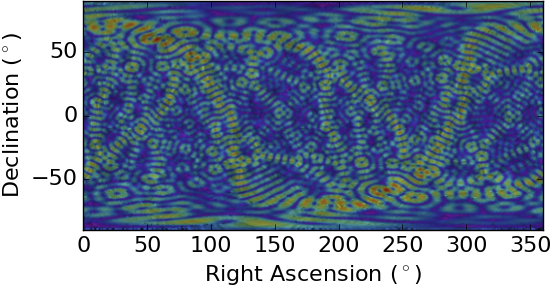}
\caption{%\footnotesize%\scriptsize
All-sky stream plots.  Streamlines indicate the vector field direction, and the colors indicate the vector amplitude, from 
violet (zero) to red (maximum).  
Top:  Secular aberration drift dipole detected by \citet{truebenbach2017}.
Middle: Randomly generated gravitational wave stream plot, after \citet{darling2018}.
Bottom:  Randomly generated BAO streamlines.}\label{fig:streamplots}
\end{figure}

Aberration of light is caused by the finite speed of light and the motion of the observer with respect to a light source.  The resulting deflection 
of light scales as $\vec{v}/c$.  If the observer accelerates, then the aberration exhibits a secular drift, and objects appear to stream in the direction of the acceleration vector.  The solar system barycenter accelerates at roughly 0.7~cm~s$^{-1}$~yr$^{-1}$ as the Sun orbits the Galactic Center, resulting
in an apparent $\sim$5~$\mu$arcsec~yr$^{-1}$ E-mode dipole converging on the Galactic Center (Figure \ref{fig:streamplots}).  This has been detected in the proper motions of 
radio sources, first by \citet{titov2011}, using {\it a priori} knowledge of the expected effect for data trimming, and recently without priors by 
\citet{truebenbach2017}.  % using a ``permissive'' fit method that allows for the large significant proper motions created by radio jets.  
\citet{titov2018} have further developed the VLBI-specific methodology for extracting this signal.  

\subsection{Secular Parallax}

The CMB shows a temperature dipole of 3.4 mK, which is caused by the motion of the solar system barycenter with respect to the CMB 
rest frame \citep[e.g.,][]{hinshaw2009}.  This amounts to a relative motion of $369\pm0.9$~km~s$^{-1}$ or 
78 AU~yr$^{-1}$.  This motion will 
induce a maximum secular parallax of 78~$\mu$arcsec~yr$^{-1}$~Mpc$^{-1}$:  nearby galaxies in directions perpendicular to the CMB
poles will show a reflex motion opposite our motion when compared to distant galaxies.  Precise proper motion measurements of galaxies
within $\sim$20~Mpc will allow a statistical detection of the distance-dependent E-mode dipole caused by secular parallax.  When 
referenced to this average dipole, it will be possible to detect the peculiar motions of individual galaxies as well as their geometric 
distances.  Secular parallax may provide a distance ladder-free method for measuring distances in the local universe.  

\subsection{Rotation}

As observers in a rotating reference frame, we have traditionally used quasars to define a fixed, non-rotating reference frame in order to 
measure and monitor the rotation of the Earth \citep[e.g.,][]{mccarthy2004}.  We thus have no means to detect non-terrestrial rotation that has an axis close to polar, but
it is possible to detect or constrain rotation about axes that are not aligned with the polar axis.  While the question of a rotating universe 
can be nettlesome to contemplate and violates our assumption of cosmological isotropy, one can nonetheless make precise, 
sub-$\mu$arcsec~yr$^{-1}$ measurements of the effect because it would manifest as a B-mode dipole in the proper motion vector field.

\subsection{Anisotropic Expansion}

In an isotropically expanding universe, objects move radially away from every observer (modulo small peculiar motions due to 
local density perturbations; see Section \ref{sec:lss}), so there will be no global correlated proper motions caused by 
isotropic Hubble expansion.  However, anisotropic expansion would cause objects to stream across the sky toward the directions of 
fastest expansion and away from directions of slowest expansion.  Assuming a simple triaxial anisotropy, one can show that the 
resulting celestial proper motion pattern can be completely described by an E-mode VSH quadrupole \citep{darling2014}.  Fitting a
curl-free quadrupole to a proper motion field can therefore measure or constrain the (an)isotropy of the Hubble expansion without
{\it a priori} knowledge of $H_0$.  We can express the Hubble constant in terms of an angular rate, $H_0 \simeq 15$~$\mu$arcsec~yr$^{-1}$,
which means that a 10\% anisotropy would produce a quadrupole with amplitude 1.5~$\mu$arcsec~yr$^{-1}$.

\subsection{Gravitational Waves}

Stochastic gravitational waves deflect light rays in a quadrupolar (and higher $\ell$) pattern with equal power in the E- and B-modes
\citep[Figure \ref{fig:streamplots};][]{pyne1996,gwinn1997,book2011}.  
The gravitational waves that will produce extragalactic proper motions lie in the frequency range
$10^{-18}$~Hz~$< f < 10^{-8}$~Hz ($H_0$ to 0.3 yr$^{-1}$), which overlaps the pulsar timing and CMB polarization regimes, but uniquely covers
about seven orders of magnitude of frequency space between the two methods \citep{darling2018}.  The cosmic energy density of gravitational
waves $\Omega_{GW}$ can be related to the proper motion variance as
\begin{equation} 
  \Omega_{\rm GW} \sim \langle\mu^2\rangle/H_0^2
\end{equation}  
and to the quadrupolar power $P_2$ as
\begin{equation} 
    \Omega_{\rm GW} = {6\over5}\,{1\over4\pi}\,{P_2\over H_\circ^2} 
                              = 0.00042\, {P_2\over (1\ \mu\rm{as\ yr}^{-1})^2}\,h_{70}^{-2}
\end{equation} 
\citep{gwinn1997,book2011,darling2018}.  
Measuring or constraining the proper motion quadrupole power 
can therefore detect or place limits on primordial gravitational waves in a unique portion of the gravitational wave spectrum.

\subsection{Large Scale Structure}\label{sec:lss}
The mass density distribution of large-scale structure reflects the mass power spectrum, the shape and evolution of which relies on cosmological parameters. The transverse peculiar motions of extragalactic objects can be used to measure the density perturbations from large-scale structure without a reliance on precise distance measurements. While line-of-sight velocity studies use distances to differentiate Hubble expansion from peculiar velocity, peculiar motions across the line-of-sight are separable from Hubble expansion because no proper motion will occur in a homogeneous expansion \citep{nusser2012,darling2013}.
Thus, one can employ pairs of galaxies as ``cosmic rulers'' to measure the real-time change in the apparent size of the rulers caused by the cosmic expansion and to 
detect structures that have decoupled from the Hubble flow \citep{darling2013}.  

Given the definition of angular diameter distance, $\theta = \ell/D_A$, where a ``ruler'' of proper length $\ell$ subtends
small angle $\theta$ at angular diameter distance $D_A$, 
cosmic expansion and a changing $\ell$ can produce an 
observed fractional rate of change in $\theta$:
\begin{equation}
{\Delta\theta/\Delta t_\circ\over\sin\theta} \equiv {\dot{\theta}\over\sin\theta} = { -\dot{D_A}\over D_A} + {\dot{\ell}\over\ell}
= {- H(z)\over1+z} + {\dot{\ell}\over\ell}
\label{eqn:pm}
\end{equation}
where $H(z) = H_\circ \sqrt{\Omega_{M,\circ}(1+z)^3+\Omega_\Lambda}$ in a flat universe, $\Delta t_\circ$ is the 
observer's time increment, $\dot{\theta}$ is the {\it relative} proper motion, and $\dot{\ell}$ is the observed 
change in proper length, $\Delta\ell/\Delta t_\circ$, related to the physical (rest-frame) 
transverse velocity as $v_\perp=\dot{\ell}\, (1+z)$.

If $\ell$ is not a gravitationally influenced structure and grows with the expansion, then
$\dot{\ell}/\ell = H(z)/1+z$, 
exactly canceling the first term in Eqn.\  (\ref{eqn:pm}).  In this case, $\dot{\theta} = 0$, and there
is no proper motion for objects co-moving with an isotropically expanding universe, as expected.  
If $\ell$ is decoupled from the expansion, however, then for most reasonable gravitational motions, 
$\dot{\ell}/\ell$ is a minor modification to the expansion contribution to $\dot{\theta}/\theta$
because the expansion, except for small redshifts or small structures, dominates \citep{darling2013}. The deviation of $\dot{\theta}/\theta$ from the null signal of pure Hubble expansion can be used to probe the mass distribution of large-scale structure and, thus, to test the shape of the mass power spectrum without a dependence on precise distance measurements or a ``distance ladder.''

\subsection{Baryon Acoustic Oscillation Evolution}

The baryon acoustic oscillation (BAO) is a standard ruler arising from pre-CMB density fluctuations, which 
can be observed as an overdensity of galaxies on the scale of $\sim$150 comoving Mpc \citep{eisenstein2005}.
At redshift $z=0.5$, the BAO scale subtends $\theta_{BAO} =  4.5^\circ$, which is equivalent to VSH degree $\ell\sim40$.  
Taking the time derivative, we obtain an expression for proper motion on these scales:
\begin{equation}
\mu_{BAO} = {\Delta\theta_{BAO}\over\Delta t_\circ} \simeq - \theta_{BAO}\ H_0 \simeq -1.2\ \mu{\rm as}\ {\rm yr}^{-1}.
\label{eqn:bao}
\end{equation}
The BAO evolution will therefore manifest as a convergent E-mode signal around $\ell\sim40$ (Figure \ref{fig:streamplots}).
To first order, the BAO scale depends on the expansion rate $H(z)$ and the angular diameter distance $D_A(z)$ at the observed
redshift, but the rate of change of this standard ruler is dominated by its recession (``receding objects appear to shrink'') and 
depends to first order on $H_0/D_A(z)$.  Detection of this effect relies critically on the sky density of sources, which must adequately sample
angular scales smaller than $4.5^\circ$.

%\begin{table}[!ht]
\begin{sidewaystable}
\caption{Global Correlated Proper Motion Signals}
\smallskip
\begin{center}
{\small
\begin{tabular}{lrccccccc}  % l = left, c = centered
\tableline
\noalign{\smallskip}
Effect & $\ell$ & Mode & Amplitude & Recent Measurement & Ref & {\it Gaia} & ngVLA & Prev.\ Work \\
\noalign{\smallskip}
 & & & ($\mu$arcsec yr$^{-1}$)  & ($\mu$arcsec yr$^{-1}$)  & & (Predicted) & (Predicted)\\
\noalign{\smallskip}
\tableline
\noalign{\smallskip}
Secular Aberration Drift & 1 &  E  &  $\sim$5  & $5.2\pm0.2$ & 1 & 10$\sigma$ & 50$\sigma$ & 2,3,4,5 \\
Secular Parallax                  & 1 & E  &   78~Mpc$^{-1}$  & ... & ... & $\sim$10$\sigma$ & $\sim$10$\sigma$~$^{(1)}$ & 6 \\
Rotation                             & 1 & B &    Unknown & $0.45\pm0.27$ & 7 & $<0.5$~$\mu$as yr$^{-1}$ & $<0.1$~$\mu$as yr$^{-1}$~$^{(2)}$ & 2,5,7 \\
Anisotropic Expansion       & 2  & E   &  Unknown & $<7$\% & 8 &  $<3$\% & $<0.7$\% & 8,9,10,11\\
Gravitational Waves           &  $\geq2$ & E+B & Unknown & $\Omega_{GW} < 6.4\times10^{-3}$ & 7 & $\Omega_{GW} < 4\times10^{-4}$& $\Omega_{GW} < 10^{-5}$ & 1,7,12,13,14,15,16 \\
Large Scale Structure          &      $\gtrsim 5$ & E & $-15$ to +5  & $8.3\pm14.9$ & 17 & 10$\sigma$ & $\sim$20$\sigma$~$^{(3)}$ & 17,18,19 \\
BAO Evolution                 & $\sim$40 & E  &  $-1.2$  at $z=0.5$  & ... & ... & 4$\sigma$ & $\sim$10$\sigma$~$^{(4)}$ & ... \\
\noalign{\smallskip}
\tableline % Sometimes you just need a line between table rows
\noalign{\smallskip}
%\noalign{\smallskip}
%\tableline\
\end{tabular}
}
\end{center}
{\small
Notes:
1 -- Detection of the secular parallax is limited by the number of compact radio sources that can (and would) be 
monitored in the local volume.
2 -- ngVLA observations will only be sensitive to rotation axes that are not aligned with the Earth's rotation axis.
3 -- Detection of peculiar velocities associated with large scale structure will depend on the number of close pairs of radio sources.
4 -- Detection of the BAO evolution signal will depend strongly on the sky density of proper motion measurements.\\ 
References:
1 -- \citet{titov2018}
2 - \citet{titov2011};  
3 -- \citet{xu2012};
4 -- \citet{titov2013};
5 -- \citet{truebenbach2017};
6 -- \citet{ding2009};
7 -- \citet{darling2018}; 
8 -- \citet{darling2014};
9 -- \citet{chang2015};
10 -- \citet{bengaly2016}; 
11 -- \citet{paine2018};
12 -- \citet{braginsky1990};
13 -- \citet{pyne1996};
14 -- \citet{kaiser1997};
15 -- \citet{gwinn1997};
16 -- \citet{book2011};
17 -- \citet{darling2013};
18 -- \citet{darling2018b};
19 -- \citet{truebenbach2018}.}
\label{Tab:Summary}
%\end{table}
\end{sidewaystable}

\section{ngVLA Science}\label{sec:ngvla}

The ngVLA can detect or constrain all of these phenomena and improve upon many of the expected {\it Gaia} measurements.
Table \ref{Tab:Summary} lists the various proper motion signals, their VSH modes, the expected amplitude of the 
signal (if known), a recent measurement (if any), and predictions for {\it Gaia} and the ngVLA.  
We make the following assumptions about a ngVLA astrometry program:
%\begin{itemize}%[noitemsep]
\begin{list}{$\bullet$}{\setlength{\topsep}{10pt} \setlength{\itemsep}{0.05in}
		\setlength{\parsep}{0pt} \setlength{\parskip}{0pt}}
 \item A sample of 10,000 objects
 \item  Astrometric observations spanning 10 years 
 \item  VLBA-level astrometric precision:  $\pm10$~$\mu$as~yr$^{-1}$ per object
%\end{itemize}
\end{list}
Implicit in these assumptions are VLBA-sized baselines with roughly ten times the current VLBA collecting area.  
We further assume that radio jets will pose the same challenges as are found in the current VLBA data.  In practice, 
we simply scale current VLBA-based observations --- which include the added intrinsic proper motion ``noise'' contribution from 
relativistic jets --- by $N$ or $\sqrt{N}$, as appropriate, to the expected ngVLA sample.  
The proposed sample is roughly ten times larger than current VLBA geodetic monitoring samples, but an increased collecting area
would enable the ngVLA (or a long-baseline subarray) to monitor the expanded sample without increasing the observing 
time commitment.  The ngVLA sample would leverage the current VLBA sample and the radio-loud {\it Gaia} AGN, providing super-decade
time baselines for a significant subset of objects.  

The proposed observations will enable global detection of correlated signals of $\sim$0.1~$\mu$as~yr$^{-1}$,
which is $\sim$0.7\% of $H_0$.  For most of the phenomena described in Section~\ref{sec:signals}, the ngVLA would
significantly improve on previous work, including the expected {\it Gaia} performance.  
While the ngVLA sample size will be a factor of $\sim$50 smaller than the 
{\it Gaia} sample, the per-source astrometry will be a factor of $\sim$20--50 times better.
We therefore predict that ngVLA observations will substantially improve on {\it Gaia} global correlated proper motion measurements
in most cases.
The proposed ngVLA program may not perform as well as our expectations for {\it Gaia} for measurements requiring fine angular 
sampling (BAO evolution) or dense volumetric sampling (secular parallax).  This is due to 
the overall physical paucity of compact radio sources compared to optical sources across the sky and in the very nearby universe.

\section{Conclusions}

The outlook for extragalactic proper motions using the ngVLA is promising, and in most cases it can 
surpass the expected performance of {\it Gaia}, despite the added challenge of the intrinsic proper 
motion caused by radio jets.    In particular, the proposed ngVLA astrometry program would provide 
exquisite precision on the Solar motion in the Galaxy and place the strongest constraints to date
on the isotropy of the Hubble expansion in the epoch of dark energy and on the primordial gravitational
wave background over roughly 10 decades in frequency.

\acknowledgements The authors acknowledge support from the NSF grant AST-1411605
and the NASA grant 14-ATP14-0086.
  % Keep this text on the same line as the \verb"\acknowledgements" command because it makes things a lot easier.

%\bibliography{editor}  % For BibTex

% For non-BibTex:

\end{document}